\documentclass[conference,a4paper]{IEEEtran}
\onecolumn
\usepackage[utf8]{inputenc} 
\usepackage[T1]{fontenc}
\usepackage{url}              
\usepackage{cite}             

\usepackage[cmex10]{amsmath}  
\usepackage{mleftright}       
\mleftright                   

\usepackage{graphicx}         
\usepackage{booktabs}         

\usepackage[utf8]{inputenc}
\usepackage[utf8]{inputenc} 
\usepackage[T1]{fontenc}
\usepackage{url}
\usepackage{ifthen}
\usepackage{cite}
\usepackage{bbm, dsfont}
\usepackage[cmex10]{amsmath} 

\usepackage{cite}
\usepackage{amsmath,amssymb,amsfonts}
\usepackage{mathtools}
\usepackage{amsthm}
\usepackage{algorithmic}
\usepackage{graphicx}
\usepackage{textcomp}
\usepackage{tikz}
\usepackage{caption}
\usepackage{cuted}
\usepackage{romannum}
\usepackage[utf8]{inputenc}
\usepackage{pgfplots} 
\usepackage{pgfgantt}
\usepackage{pdflscape}
\usepackage{amssymb}
\usepackage{comment}
\usepackage{pst-plot}
\usetikzlibrary{spy}
\usetikzlibrary{positioning,calc}
\usetikzlibrary{decorations.pathmorphing,calc,shapes,shapes.geometric,patterns}
\usetikzlibrary{shapes.multipart}
\usepackage{xfrac}
\usepackage{colortbl}
\usepackage{cancel}
\usetikzlibrary{arrows,positioning,calc,intersections}
\usetikzlibrary{datavisualization.formats.functions}
\def\BibTeX{{\rm B\kern-.05em{\sc i\kern-.025em b}\kern-.08em
    T\kern-.1667em\lower.7ex\hbox{E}\kern-.125emX}}
    
\usepackage{pgfplots}
\usepgfplotslibrary{fillbetween}
\usetikzlibrary{arrows, decorations.markings}
\newtheorem{theorem}{Theorem}
\newtheorem{lemma}{Lemma}

\newtheorem{definition}{Definition}

\newcommand{\bs}[1]{\boldsymbol{#1}}

\definecolor{calpolypomonagreen}{rgb}{0.12, 0.3, 0.17}
\newcounter{remarkcount}
\newenvironment{remark}{\refstepcounter{remarkcount}\begin{trivlist}\item \textbf{Remark \theremarkcount.}}{\end{trivlist}}
\newcommand{\circlearrow}{}
\DeclareRobustCommand{\circlearrow}{%
  \mathrel{\vphantom{\rightarrow}\mathpalette\circle@arrow\relax}%
}
\newcommand{\circle@arrow}[2]{%
  \m@th
  \ooalign{%
    \hidewidth$#1\circ\mkern1mu$\hidewidth\cr
    $#1-$\cr}%
}
\makeatother

\usepackage{amsmath, amssymb, amsfonts, amsthm}
\usepackage{bm}
\usepackage{xcolor}
\usepackage{framed}

\usepackage{pgf}
\usepackage{tikz}
\usetikzlibrary{arrows,automata}
\usetikzlibrary{positioning}
\usetikzlibrary{decorations.pathmorphing}
\usetikzlibrary{shapes.geometric}
\usetikzlibrary{fit}
\usetikzlibrary{backgrounds}

\usepackage[caption=false,font=footnotesize]{subfig}

\newcommand{\mbf}{\mathbf}
\newcommand{\mc}{\mathcal}
\newcommand{\mbb}{\mathbb}

\newtheorem{corollary}{Corollary}

\theoremstyle{definition}

\theoremstyle{remark}

\def\BibTeX{{\rm B\kern-.05em{\sc i\kern-.025em b}\kern-.08em
    T\kern-.1667em\lower.7ex\hbox{E}\kern-.125emX}}
\begin{document}

\title{Common Randomness Generation from Finite Compound Sources} 

\author{
\IEEEauthorblockN{Rami Ezzine\IEEEauthorrefmark{1}\IEEEauthorrefmark{4}, Moritz Wiese\IEEEauthorrefmark{1}\IEEEauthorrefmark{4}, Christian Deppe\IEEEauthorrefmark{2}\IEEEauthorrefmark{4}\IEEEauthorrefmark{6} and Holger Boche\IEEEauthorrefmark{1}\IEEEauthorrefmark{3}\IEEEauthorrefmark{4}\IEEEauthorrefmark{5}\IEEEauthorrefmark{6}}
\IEEEauthorblockA{\IEEEauthorrefmark{1}Technical University of Munich, Munich, Germany\\
\IEEEauthorrefmark{2}Technical University of Braunschweig, Brunswick, Germany\\
\IEEEauthorrefmark{3}CASA -- Cyber Security in the Age of Large-Scale Adversaries–
Exzellenzcluster, Ruhr-Universit\"at Bochum, Germany\\
\IEEEauthorrefmark{4}BMBF Research Hub 6G-life, Munich, Germany\\
\IEEEauthorrefmark{5}{\color{black}{ Munich Center for Quantum Science and Technology (MCQST) }}\\
\IEEEauthorrefmark{6} {\color{black}{Munich Quantum Valley (MQV)}} \\
Email: \{rami.ezzine@tum.de, wiese@tum.de, christian.deppe@tu-braunschweig.de, boche@tum.de\}}
}\color{black}

\maketitle
\thispagestyle{plain}
\pagenumbering{arabic}
\pagestyle{plain}
\begin{abstract}
We investigate the problem of generating common randomness (CR) from finite compound sources aided by unidirectional communication over rate-limited perfect channels. The two communicating parties, often referred to as terminals, observe independent and identically distributed (i.i.d.) samples of a finite compound source and aim to agree on a common random variable with a high probability for every possible realization of the source state. Both parties know the set of source states as well as their statistics. However, they are unaware of the actual realization of the source state. We establish a single-letter lower and upper bound on the compound CR capacity for the specified model. Furthermore, we present two special scenarios where the established bounds coincide.
\end{abstract}

\section{Introduction}
\color{black} In the problem of common randomness (CR) generation, \color{black} communicating parties aim to establish a common random variable with a high probability\cite{part2}. 

\color{black}CR is viewed as a highly promising resource for future communication systems.\color{black} \ For instance, CR brings significant performance gains in Post-Shannon communication tasks, such as identification and secure identification. Indeed, this resource can significantly increase the identification/secure identification capacity of channels \cite{Generaltheory}. While the number of identification messages (also called identities) increases exponentially with the block length in the deterministic identification scheme for discrete memoryless channels (DMCs), the size of the identification code increases doubly exponentially with the block length when CR is used as a resource. The identification scheme \cite{identification}, developed by Ahlswede and Dueck in 1989, is more efficient than Shannon's classical transmission scheme\cite{shannon} in various fields such as machine-to-machine and human-to-machine systems\cite{applications}, industry 4.0 \cite{industrie40}, and 6G communication systems\cite{6Gcomm}\cite{6Gpostshannon}. The identification scheme's applications extend to digital watermarking\cite{Moulin,watermarkingahlswede,watermarking}.

The Post-Shannon resource of CR can also be exploited to achieve resilience by design for the tactile internet and quantum communication systems. Indeed, when legitimate parties have access to a common random source as an additional resource for coordination, communication becomes robust against denial-of-service (DOS) attacks from the jammer, and only a few bits of CR are required to nullify the jamming attack\cite{jammerref}.

Moreover, CR is highly relevant in the modular coding scheme for secure communication. Modular schemes for semantic security have been developed in \cite{semanticwiretap}. \color{black} They allow to establish semantic security in combination with arbitrary error-correcting codes. \color{black} A common scenario in seeded modular coding is when legitimate parties have access to CR as an additional resource. This CR can be used as a seed \cite{semanticsecurity}.

Furthermore, CR is of high relevance in the key generation problem from correlated observations. Under additional secrecy constraints, the generated CR can be used in the information reconciliation step in secret key generation, as shown in the fundamental two papers \cite{part1,Maurer}. The generated secret keys are used to perform cryptographic tasks, including secure message transmission and message authentication. 

Various information-theoretical models for generating CR have been explored in the literature \cite{part2,survey,helper,crpair}. One of the most commonly studied models is the two-source model with one-way communication, originally introduced by Ahlswede and Csiszár in \cite{part2}. In this model, the two terminals observe independent and identically distributed (i.i.d.) samples from a correlated finite source. Ahlswede and Csiszár considered the case when the two terminals communicate over perfect channels with limited capacity, as well as the case when the terminals communicate over DMCs. They derived a single-letter formula for the common randomness (CR) capacity in both scenarios.
Later, the problem of CR generation from finite sources with unidirectional communication over Gaussian channels and \color{black}over slow fading channels with arbitrary state distribution \color{black} has been investigated in \cite{CRgaussian} and \cite{CRfading}, respectively. \color{black}Subsequently, \cite{UCRgeneral} and \cite{EpsilonUCRgeneral} studied the problem of uniform common randomness generation from finite sources aided by a one-way communication over arbitrary point-to-point channels. \color{black}

Nevertheless, to the best of our knowledge, the problem of CR generation from compound sources has not been addressed in the literature so far.
The concept of generating shared randomness from compound sources was introduced solely in the context of secret key generation\cite{compound3,compound4,compoundsecretkey,authentication,compound1,compound2}, where additional requirements regarding secrecy are imposed.

In our work, no such secrecy constraints are imposed. We consider a two-source model for CR generation from finite compound sources with one-way communication over perfect channels with limited capacity. 
Compound sources illustrate a practical and realistic scenario concerning source state information, where legitimate users lack precise awareness of the actual source realization. However, they possess knowledge that the source derives from a defined uncertainty set, and the latter remains unchanged during the observation time scale. We are specifically interested in a scenario where only the source outputs observed by \color{black}the terminal at the receiving end of the noiseless channel \color{black} are state-dependent.

The main contribution of our work consists in establishing single-letter lower and upper bound on the \textit{compound} \ CR capacity for the proposed model.
 
\textit{Outline:} The rest of the paper is organized as follows. In Section \ref{sec2}, we present our system model for CR generation from finite compound source with one-way communication over rate-limited perfect channels, review the definition of an achievable \textit{compound} CR rate and the \textit{compound} \ CR capacity and present our main result. In Section \ref{prooflowerbound}, we prove the lower bound on the \textit{compound} \ CR capacity. Section \ref{proofupperbound} is dedicated to the proof of the upper-bound on the \textit{compound} CR capacity. In Section \ref{specialcase1} and \ref{specialcase2}, we consider two special scenarios, where the established bounds coincide. Section \ref{conclusion} contains concluding remarks. The appendix contains a technical proof.

\textit{Notation:} Throughout the paper,  $\log$ is taken to  base 2 and $\ln$ refers to the natural logarithm. For any set $\mc E,$ $\mc E^c$ refers to its complement and $\lvert \mc E \rvert$ refers to its cardinality.  For any random variable $X$ with distribution $P_X,$ $\text{supp}(P_X)$ refers to its support.
Let $\sigma>0$ be fixed arbitrarily and let $X,$ $Y$ and $Z$ be any discrete random variables with joint distribution $P_{X,Y,Z}$ and support $\mc X\times \mc Y \times \mc Z.$ We define
\begin{align}
    \mc T_{\sigma}^{n}(P_X)=\bigg\{ x^n \in \mc X^n: \bigg| \frac{N(a|x^n)}{n}-P_{X}(a) \bigg| \leq \sigma P_{X}(a) \quad  \forall a \in \mc X\bigg\}, \nonumber
\end{align}
where $N(a|x^n)$ refers to the number of positions of $x^n$ having the letter $a.$
Furthermore,  we define
\begin{align}
   \mc T_{\sigma}^{n}(P_{X,Y})
    =\bigg\{ (x^n,y^n) \in \mc X^n\times \mc Y^n: \bigg| \frac{N(a,b|x^n,y^n)}{n}-P_{X,Y}(a,b) \bigg| \leq \sigma P_{X,Y}(a,b) \quad  \forall (a,b) \in \mc X \times \mc Y \bigg\}, \nonumber
\end{align}
where $N(a,b|x^n,y^n)$ refers to the number of positions of $(x^n,y^n)$ having the pair  $(a,b),$
and 
\begin{align}
    \mc T_{\sigma}^{n}(P_{X,Y,Z}) 
    &=\bigg\{ (x^n,y^n,z^n) \in \mc X^n\times \mc Y^n \times \mc Z^n:   \bigg| \frac{N(a,b,c|x^n,y^n,z^n)}{n}-P_{X,Y,Z}(a,b,c) \bigg| \leq \sigma P_{X,Y,Z}(a,b,c)  \nonumber \\ &\quad \quad \quad \forall (a,b,c) \in \mc X \times \mc Y \times \mc Z\bigg\}, \nonumber
\end{align}
where $N(a,b,c|x^n,y^n,z^n)$ refers to the number of positions of $(x^n,y^n,z^n)$ having the triple $(a,b,c).$

\section{System Model, Definitions and Main Result}
\label{sec2}
\subsection{System Model}
\label{systemmodel}
\begin{figure}[h!]
\centering
\tikzstyle{block} = [draw, rectangle, rounded corners,
minimum height=2em, minimum width=2cm]
\tikzstyle{blockchannel} = [draw, top color=white, bottom color=white!80!gray, rectangle, rounded corners,
minimum height=1cm, minimum width=.3cm]
\tikzstyle{input} = [coordinate]
\usetikzlibrary{arrows}
\begin{tikzpicture}[scale= 1,font=\footnotesize]
\node[blockchannel] (source) { Compound DMMS $\{XY_{s}\}_{s\in \mc S}$};
\node[blockchannel, below=2.6cm of source](channel) { Noiseless Rate-limited Channel};
\node[block, below left=2.2cm of source] (x) {Terminal $A$};
\node[block, below right=2cm of source] (y) {Terminal $B$};
\node[above=1cm of x] (k) {$K=\Phi(X^n)$};
\node[above=1cm of y] (l) {$L_{s}=\Psi(Y_{s}^n,f(X^n))$};

\draw[->,thick] (source) -- node[above] {$X^n$} (x);
\draw[->, thick] (source) -- node[above] {$Y_{s}^n$} (y);
\draw [->, thick] (x) |- node[below right] {$f(X^n)$} (channel);
\draw[<-, thick] (y) |- node[below left] {$f(X^n)$} (channel);
\draw[->] (x) -- (k);
\draw[->] (y) -- (l);

\end{tikzpicture}
\caption{Two-source model for CR generation from finite compound sources with one-way communication over rate-limited perfect channels.}
\label{CRprotocol}
\end{figure}
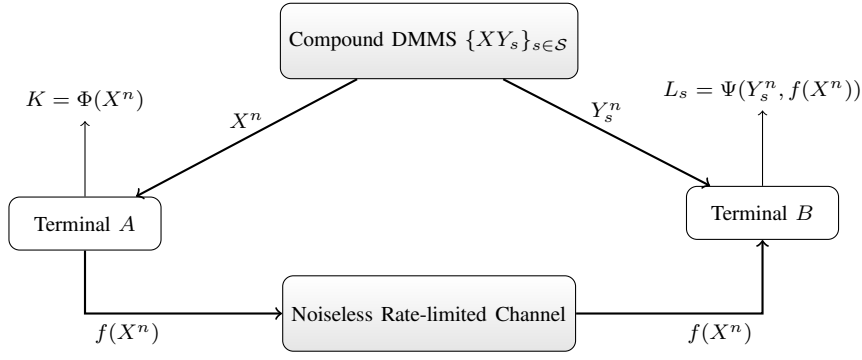

Let a compound discrete memoryless multiple source (CDMMS) $\{ XY_{s}   \}_{s\in\mc S}$ on alphabets $\mathcal{X}$ and $\mathcal{Y}$, respectively, be given, \color{black}where the set of source states $\mc S$ is finite. \color{black} 
The  CDMMS emits $n$ i.i.d. samples of $(X,Y_{s}).$
Suppose that the outputs of $X$ are observed only by Terminal $A$ and those of $Y_s$ only by Terminal $B.$ \color{black} We assume that $X^n$ and all the ${Y_{s}^{n}}'s$ are defined on a joint probability space. \color{black} Assume also that the joint distribution of $(X, Y_s)$ is known to both terminals for all $s\in\mc S.$ Additionally, both terminals know the set of source states  as well as their statistics with probability distributions $\{ 
P_{XY_{s}}\}_{s\in \mc S}.$  However, they don't know \color{black}the actual state $s\in \mc S.$ \color{black}    
Terminal $A$
can communicate with Terminal $B$ over a noiseless channel with limited capacity $R>0.$
There are no other resources available to any of the terminals. 
\begin{definition}
\label{CRprotocoldef}
A CR-generation protocol of block-length $n$ consists of:
\begin{enumerate}
    \item A function $\Phi$ \color{black} at Terminal $A$ which \color{black} maps $X^n$ into a random variable $K$ with alphabet $\mathcal{K}.$
    \item A function $f$ \color{black} at Terminal $A$ which \color{black} maps $X^n$ into a message $f(X^n)$ such that  \footnote{$\lVert f \rVert$ 
 \text{refers to the number of messages. This is the same notation used in} \cite{codingtheorems}.}  $\frac{\log \lVert f \rVert}{n}\leq R,$ with $R>0.$
    \item A function $\Psi$ \color{black} at Terminal $B$ which \color{black}  maps $Y_s^n$ and $f(X^n)$ into a random variable $L_s$ with alphabet $\mathcal{K}.$ 
\end{enumerate}
\color{black}Such a protocol induces for every $s\in \mc S$ a pair of random variables $(K,L_s)$, where the distribution of $K$ is independent of the state $s.$ This pair of random variables is called permissible.
This is illustrated in Fig. \ref{CRprotocol}. \color{black}
\end{definition}
\subsection{Achievable Rate and Capacity}
We define first an achievable \textit{compound} CR rate and the \textit{compound} CR capacity.
\begin{definition} 
\label{ucrrate}
 A number $H$ is called an achievable compound CR rate  if there exists a non-negative constant $c$ such that for every $\alpha,\delta>0$  and for sufficiently large $n$ there exists a permissible  pair of random variables $(K,L_s)$ for every $s\in \mc S$ such that
\begin{equation}
  \forall s \in \mathcal{S}: \mbb P\left[K\neq L_s\right]\leq \alpha, 
    \label{errorcorrelated}
\end{equation}
\begin{equation}
    |\mathcal{K}|\leq 2^{cn},
    \label{cardinalitycorrelated}
\end{equation}
\begin{equation}
    \frac{1}{n}H(K)> H-\delta.
     \label{ratecorrelated}
\end{equation}
\end{definition}
\begin{definition} 
The compound CR capacity $C_{CCR}(R)$ is the maximum achievable compound CR rate.
\end{definition}
\subsection{Main Result}
In this section, we give an upper and a lower bound on the CCR capacity for the model presented in Section \ref{systemmodel}.
\begin{theorem}
\label{compound CRcapacity}
For the model in Fig \ref{CRprotocol}, the \textit{compound} CR capacity $C_{CCR}(R)$ satisfies
\begin{align}
C_{CCR}(R)\geq \underset{ \substack{U \\{\substack{\forall s \in \mc S: U\circlearrow{X} \circlearrow{Y_s}\\ I(U;X)-\underset{s\in \mc S}{\min}I(U;Y_s) \leq R}}}}{\max} I(U;X)
\label{lowerboundepsilonCRcapacity}
\end{align}
and
\begin{align}
C_{CCR}(R)\leq \underset{s\in \mc S}{\min} \underset{ \substack{U_s \\{\substack{U_s \circlearrow{X} \circlearrow{Y_s}\\ I(U_s;X)-I(U_s;Y_s) \leq R}}}}{\max} I(U_s;X).
\label{upperboundepsilonCRcapacity}
\end{align}
\end{theorem}
\begin{corollary}
\label{corollary1}
  \color{black} For $R\geq \underset{s\in \mc S}  
  {\max}\ H(X|Y_{s}),$ the bounds in \eqref{lowerboundepsilonCRcapacity} and \eqref{upperboundepsilonCRcapacity} coincide, and they are both equal to $H(X).$ 
\end{corollary}
\begin{remark}
   For $R<\underset{s\in \mc S}  
  {\max}\ H(X|Y_{s}),$ the right-hand side of $\eqref{lowerboundepsilonCRcapacity}$ is lower-bounded by $R$ since the constraints in the right-hand side of \eqref{lowerboundepsilonCRcapacity} are satisfied for $U=g(X),$ with $H(g(X))=R<\underset{s\in \mc S}  
  {\max}\ H(X|Y_{s})\leq H(X).$ This lower-bound is not tight. To see this, consider the case when $\mc S=\{s_0\}$  such that $X=Y_{s_{0}}.$ Then, in this case, the right-hand side of \eqref{lowerboundepsilonCRcapacity} is equal to $H(X).$
\end{remark}
\begin{corollary}
\label{corollary2}
    If there exists $s'\in \mc S$ such that for every $s\in \mc S,$ $X\circlearrow{Y_{s}}\circlearrow{Y_{s^\prime}}$ forms a Markov chain, \color{black}then, \color{black} the bounds in \eqref{lowerboundepsilonCRcapacity} and in \eqref{upperboundepsilonCRcapacity} are equal.
\end{corollary}

\color{black}Notice that the function $R\mapsto C_{CCR}(R)$ is non-decreasing in $(0,\infty).$ This is clear from Definition \ref{CRprotocoldef}, since increasing the channel capacity cannot decrease the compound CR capacity.\color{black}

\section{Proof of the Lower-bound in Theorem \ref{compound CRcapacity} }
\label{prooflowerbound}
We extend the CR generation scheme provided in \cite{part2} to finite compound sources. By continuity, it suffices to show that 
$$ \underset{ \substack{U \\{\substack{U \circlearrow{X} \circlearrow{Y_s}\\ I(U;X)- \underset{s\in \mc S}{\min} \ I(U;Y_s) \leq R'}}}}{\max} I(U;X)  $$ is an achievable \color{black} compound \color{black} CR rate for every $R'<R.$
Let $U$ be any random variable with alphabet $\mc U$ satisfying for every $s\in \mathcal{S}:$ $U \circlearrow{X} \circlearrow{Y_s}$ and $I(U;X)-\underset{s\in \mc S}{\min}I(U;Y_{s}) \leq R'$. We are going to show that $H=I(U;X)$ is an achievable \color{black} compound \color{black} CR rate.  Let $\alpha,\delta>0$ be fixed arbitrarily.  
For some $\sigma>0,$ let $\sigma_1=\sigma,$ $\sigma_2=2\sigma$ and $\sigma_3=3\sigma.$ Let also $4\sigma_2 H(U)<\mu<5\sigma_2 H(U).$
Next, define
{{\begin{align}
N_{1}&=\lfloor 2^{n[I(U;X)-\underset{s\in \mc S}{\min}I(U;Y_s)+3\mu]} \rfloor \nonumber\\
N_{2}&=\lfloor 2^{n[\underset{s\in \mc S}{\min}I(U;Y_s)-2\mu]}\rfloor. \nonumber
\end{align}}} \color{black} On the same probability space on which $X^n$ and all the ${Y_{s}^{n}}'s$ are defined, we define $N_1 N_2$ random sequence $\bs{U}_{i,j}\in\mathcal{U}^n, i=1,\hdots,N_{1}, j=1,\hdots,N_{2},$  where \color{black} each $\bs{U}_{i,j}$ is uniformly distributed on the set $\mc T_{\sigma_1}^{n}(P_{U}).$   \color{black}We assume that the $\bs{U}_{i,j}s$ are independent. \color{black} Let $\mbf M=\bs{U}_{1,1},\hdots, \bs{U}_{N_{1},N_{2}}$  be the joint random variable of all $\bs{U}_{i,j}s.$  We define $\Phi_{\mbf M}$ as follows:
 Let $\Phi_{\mbf M}(X^n)=\bs{U}_{ij}$, if 
 $(\bs{U}_{ij},X^n)  \in \mc T_{\sigma_2}^{n}(P_{UX})$ (either one if there are several). If no such $\bs{U}_{i,j}$ exists, then  $\Phi_{\mbf M}(X^n)$ is set to a constant sequence $\bs{u}_0$ different from all the  ${\bs{U}_{ij}}s$, known to both terminals and such that $(\bs{u}_0,X^n)\notin \mc T_{\sigma_2}^{n}(P_{U,X})$.  Let $f(X^n)=i$ if $\Phi(X^n)=\bs{u}_{i,j}$. Otherwise, if $\Phi(X^n)=\bs{u}_{0},$ then $f(X^n)=N_1+1.$ We further define for any $s\in \mc S$ the two sets
\begin{align}
    \mc E_{1}^{s}(\mbf M)&=\{(x^{n},y^{n})\in \mc X^n\times\mc Y^n:\ (\Phi_{\mbf M}(x^{n}),x^{n},y^{n}) \notin \mathcal{T}_{\sigma_3}^{n} (P_{U,X,Y_s})\} \nonumber
\end{align} and
\begin{align}
    &\mc E_{2}^{s}(\mbf M) \nonumber \\
    &=\Big\{(x^{n},y^{n})\in \mc X^n \times \mc Y^n: (x^{n},y^{n}) \in {\mc E_{1}^{s}(\mbf M)}^{c} \ \text{with} \ \bs{U}_{i,j}=\Phi_{\mbf M}(x^{n}) \nonumber \\   & \ \ \ \  \ \text{and such that} \ \exists \ \bs{U}_{i,\ell}\neq\bs{U}_{i,j} \ \text{where} \ (\bs{U}_{i,\ell},y^n)\in \mc \mc T_{\sigma_2}^{n}(P_{U,Y_s}) (\text{with the same first index} \ i)
\Big\}.\nonumber
\end{align}
Furthermore, define the events
\begin{align}
    \mathcal{A}_{\mbf M}= ``(X^n,Y_s^{n})\in  \mc E_{1}^{s}(\mbf M) \ \text{for all} \ s \in \mc S" \nonumber
\end{align}
and
\begin{align}
    \mathcal{B}_{\mbf M}=``\exists s\in \mc S: (X^n,Y_s^{n})\in  \mc E_{2}^{s}(\mbf M)". 
\end{align}
It is proved in the Appendix that
\begin{align}
    \color{black}\mathbb{E}_{\mbf M}\left[ \mbb P\left[ \mathcal{A}_{\mbf M}|\mbf M\right]+\mbb P\left[ \mathcal{B}_{\mbf M}|\mbf M\right]\right]\leq \zeta(n), \color{black}
    \label{averagebeta}
\end{align}
where $\underset{n\rightarrow\infty}{\lim} \zeta(n)=0.$ \color{black} It is worth-mentioning here  that the joint distribution of the ${Y_{s}^{n}}'s$ is irrelevant in the proof of \eqref{averagebeta}. \color{black} 
We choose a realization $\mbf m=\bs{u}_{1,1},\hdots, \bs{u}_{N_1,N_2}$ satisfying:
\begin{align}
\mbb P\left[\mathcal{A}_{\mbf m}\right]+\mbb P\left[\mathcal{B}_{\mbf m}\right]\leq 2\zeta(n).
\label{choiceofm} \end{align} 
From \eqref{averagebeta} and using Markov inequality, we know that such a realization exists. We denote $\Phi_{\mbf m}$ by $\Phi.$
We assume that each $\bs{u}_{i,j}, \ i=1\hdots N_1, \ j=1\hdots N_2,$  is known to both terminals.  
This means that  $N_{1}$ codebooks $C_{i}, 1\leq i \leq N_{1}$, are known to both terminals, where each codebook contains $N_{2}$ sequences, $ \bs{u}_{i,j}, \ j=1,\hdots, N_2$. 

Let $s\in \mc S$ be fixed arbitrarily. Let $x^{n}$ be any realization of $X^n$ and $y_s^{n}$ be any realization of $Y_s^n.$ 

  Since $ R'<R$, we choose $\sigma$ (and consequently $\sigma_2$ and $\mu$) to be sufficiently small such that
      \begin{align}
     \frac{\log(N_1+1)}{n} 
     &\leq R'-\mu',
     \label{inequalitylogfSISO}
      \end{align}
for some $\mu'>0.$
 The message $i^\star=f(x^{n})$, with $i^\star\in\{1,\hdots,N_1+1\}$, is sent over the noiseless channel.
Let $\Psi(y_s^{n},f(x^{n}))=\bs{u}_{i^\star,j}$ if there exists $s^\star \in \mc S$ such that $(\bs{u}_{i^\star,j},y_s^{n}) \in \mc T_{\sigma_2}^n (P_{UY_s^\star}).$   If no such $s^\star$ exists or if such $s^\star$ exists but there are several $\bs{u}_{\tilde{i}^\star,j}s$ satisfying $(\bs{u}_{\tilde{i}^\star,j},y_s^{n}) \in \mc T_{\sigma_2}^n (P_{UY_s^\star}),$ then, we set $\Psi(y_s^{n},f(x^n))=\bs{u}_0.$ 
We will show next that the following requirements are satisfied for sufficiently large $n$:
\begin{equation}
 \mbb P\left[K\neq L_s\right]\leq \alpha, 
    \label{errorcorrelated2}
\end{equation}
\begin{equation}
    |\mathcal{K}|\leq 2^{cn},
    \label{cardinalitycorrelated2}
\end{equation}
\begin{equation}
    \frac{1}{n}H(K)> H-\delta.
     \label{ratecorrelated2}
\end{equation}

We define for any $(i,j)\in \{1,\hdots,N_{1}\}\times\{1,\hdots,N_{2}\}$  the set
$$\mc R=\{ x^{n}\in\mathcal{X}^{n}: (\bs{u}_{i,j},x^{n}) \in \mc T_{\sigma_2}^{n}(P_{UX})\}.$$
Then, it holds that for any $(i,j)\in \{1,\hdots,N_{1}\}\times\{1,\hdots,N_{2}\}$ 
\begin{align}
\mbb P[K=\bs{u}_{i,j}] &\overset{(a)}{=}\sum_{x^{n}\in\mc R}\mbb P[K=\bs{u}_{i,j}|X^n=x^{n}]P_{X}^n(x^{n}) \nonumber \\
&\leq \sum_{x^{n}\in\mc R}P_{X}^n(x^{n}) \nonumber \\
&=\mbb P\left[ (X^n,\bs{u}_{i,j})\in \mc T_{\sigma_2}^{n}(P_{UX}) \right] \nonumber\\
& \leq 2^{-n\left[I(U;X)-2\sigma_2 H(U)  \right]} \nonumber
\end{align}
where $(a)$ follows because for  $(\bs{u}_{i,j},\mathbf{x}) \notin \mc T_{\sigma_2}^{n}(P_{UX}),$ we have $\mbb P[K=\bs{u}_{i,j}|X^n=x^{n}]=0.$ 

As a result, we have
\begin{align}
    H(K)&\geq \sum_{i=1}^{N_1}\sum_{j=1}^{N_2} \mbb P\left[K=\bs{u}_{i,j} \right]\log\frac{1}{\mbb P\left[ K=\bs{u}_{i,j}\right]} \nonumber \\
    &= n\left[I(U;X)-2\sigma_2 H(U)\right] \left(1-\mbb P\left[K=\bs{u}_0 \right]\right)
\nonumber \end{align}
Now, from the choice of $\mbf m$ in \eqref{choiceofm}, we know that
\begin{align}
    \mbb P\left[K=\bs{u}_0 \right]=\mbb P\left[ \cap_{i,j} (\bs{u}_{i,j}, X^n)\notin \mc T_{\sigma_2}^{n}(P_{UX}) \right] \leq \zeta'(n)
\nonumber \end{align}
for some $\zeta'(n)\leq 2 \zeta(n).$
Thus, it follows that
\begin{align}
    \frac{H(K)}{n} &\geq \left[I(U;X)-2\sigma_2 H(U)\right] \left(1-2\zeta(n)\right) \nonumber \\
    &\geq I(U;X)-2\sigma_2 H(U)-2\zeta(n)I(U;X). \nonumber 
\end{align}
Recall that $\underset{n\rightarrow \infty}{\lim}\zeta(n)=0.$  Therefore, for sufficiently large $n$ and by making $\sigma_2=2\sigma$ sufficiently small, it follows that
\begin{align}
    \frac{H(K)}{n}\geq I(U;X)-\delta.
\nonumber \end{align}
Thus, \eqref{ratecorrelated2} is satisfied.
For $c=I(U;X)+\mu+1,$  we have $\lvert \mc K \rvert = N_1 N_2+1 \leq 2^{nc}$ and therefore \eqref{cardinalitycorrelated2} is also satisfied.

Now, it remains to prove $\eqref{errorcorrelated2}.$
Let $$\mathcal{D}_{\mbf m}= ``\Phi(X^n) \ \text{is equal to none of the} \  {\bs{u}_{i,j}}'s".$$
We denote its complement by $\mc D_{\mbf m}^{c}.$
It holds that for any $s\in \mc S$
\begin{align}
    &\mbb P[K\neq L_s] \nonumber \\
    &\overset{(a)}{=}\mbb P[K\neq L_s|\mathcal{D}_{\mbf m}^c]\mbb P[\mathcal{D}_{\mbf m}^c] \nonumber \\
   &\leq \mbb P[K\neq L_s|\mathcal{D}_{\mbf m}^c],\nonumber
\end{align}
where $(a)$ follows from $\mbb P[K\neq L_s|\mathcal{D}_{\mbf m}]=0,$ since conditioned on $\mathcal{D}_{\mbf m}$, we know that $K$ and $L_s$ are both equal to $\bs{u}_0$.
It follows that
\begin{align}
    \mbb P[K\neq L_s]&\leq \mbb P[K\neq L_s|\mathcal{D}_{\mbf m}^c] \nonumber \\
    &\leq \mbb P\left[\mathcal{A}_{\mbf m}\cup \mathcal{B}_{\mbf m} \right] \nonumber \\
    &\overset{(a)}{\leq}\mbb P\left[\mathcal{A}_{\mbf m}\right]+\mbb P\left[\mathcal{B}_{\mbf m}\right]\nonumber 
\end{align}
where $(a)$ follows from the union bound. 
It follows using \eqref{choiceofm} that 
\begin{align}
        \mbb P[K\neq L_s] &\leq  2\zeta(n),\nonumber\\
    &\overset{(a)}{\leq} \alpha\nonumber 
\end{align}
where $(a)$ follows because $\zeta(n)\leq \frac{\alpha}{2}$  for sufficiently large $n$ since $\underset{n\rightarrow \infty}{\lim} \zeta(n)=0.$ This proves \eqref{errorcorrelated2}. 
This completes the proof of the lower-bound on the compound CR capacity.
\section{Proof of the upper-bound in Theorem \ref{compound CRcapacity}}
\label{proofupperbound}

Let $H$ be any achievable compound CR rate.  So, there exists a non-negative constant $c$ such that for every $\alpha,\delta>0$ and for sufficiently large $n,$ there exists for every $s\in \mc S$ a permissible pair of random variables $(K,L_s)$ according to a fixed CR-generation protocol of block-length $n,$ \color{black} where the distribution of $K$ is independent of the state $s,$ \color{black} such that 
\begin{equation}
  \forall s \in \mathcal{S}: \mbb P\left[K\neq L_s\right]\leq \alpha, 
    \label{errorcorrelated1}
\end{equation}
\begin{equation}
    |\mathcal{K}|\leq 2^{cn},
    \label{cardinalitycorrelated1}
\end{equation} and
\begin{equation}
    \frac{1}{n}H(K)> H-\delta.
     \label{ratecorrelated1}
\end{equation}

 Let $s\in\mc S$ be fixed arbitrarily. Now, let $J$ be a random variable uniformly distributed on $\{1,\dots, n\}$ and independent of $K$, $X^n$ and $Y_s^n=Y_{s,1},\hdots,Y_{s,n}$. We further define $U_s=(K,X_{1},\dots, X_{J-1},Y_{s,J+1},\dots, Y_{s,n},J).$ It holds that $U_s \circlearrow{X} \circlearrow{Y_{s,J}}.$ 
 Notice  that
 {{\begin{align}
 		H(K)&\overset{(a)}{=}H(K)-H(K|X^{n})\nonumber\\
 		&=I(K;X^{n}) \nonumber\\
 		&\overset{(b)}{=}\sum_{i=1}^{n} I(K;X_{i}|X_{1},\dots, X_{i-1}) \nonumber\\
 		&=n I(K;X_{J}|X_{1},\dots, X_{J-1},J) \nonumber\\
 		&\overset{(c)}{\leq }n I(U_s;X_{J}), \nonumber
 		\end{align}}}where $(a)$ follows because $K=\Phi(X^n)$ and $(b)$ and $(c)$ follow from the chain rule for mutual information.
\color{black} 		
 Let us now introduce the following lemma:
 \begin{lemma} (Lemma 17.12 in \cite{codingtheorems})
 	For arbitrary random variables $R_1$ and $R_2$ and sequences of random variables $X^{n}$ and $Y^{n}$, it holds that
 	\begin{align}
 	&I(R_1;X^{n}|R_2)-I(R_1;Y^{n}|R_2) \nonumber \\
 	&=\sum_{i=1}^{n} I(R_1;X_{i}|X_{1},\dots, X_{i-1}, Y_{i+1},\dots, Y_{n},R_2) \nonumber \\ &\quad -\sum_{i=1}^{n} I(R_1;Y_{i}|X_{1},\dots, X_{i-1}, Y_{i+1},\dots, Y_{n},R_2) \nonumber \\
 	&=n[I(R_1;X_{J}|V)-I(R_1;Y_{J}|V)],\nonumber
 	\end{align}
 	where $V=(X_{1},\dots, X_{J-1},Y_{J+1},\dots, Y_{n},R_2,J)$, with $J$ being a random variable independent of $R_1$,\ $R_2$, \ $X^{n}$ \ and $Y^{n}$ and uniformly distributed on $\{1 ,\dots, n \}$.
 	\label{lemma1}
 \end{lemma}
 Applying Lemma \ref{lemma1} for $R_1=K$, $R_2=\varnothing$ with $V_s=(X_1,\hdots, X_{J-1},Y_{s,J+1},\hdots, Y_{s,n},J)$ yields \color{black}
 {{\begin{align}
 		&I(K;X^{n})-I(K;Y_s^{n}) \nonumber \\
 		&=n[I(K;X_{J}|V_s)-I(K;Y_{s,J}|V_s)] \nonumber\\
 		&\overset{(a)}{=}n[I(KV_s;X_{J})-I(V_s;X_{J})-I(KV_s;Y_{s,J})+I(V_s;Y_{s,J})] \nonumber\\ 
 		&\overset{(b)}{=}n[I(U_s;X_{J})-I(U_s;Y_{s,J})], 
 		\label{UhilfsvariableMIMO1}
 		\end{align}}}where $(a)$ follows from the chain rule for mutual information and from the fact that $V_s$ is independent of $(X_{J},Y_{s,J})$ and $(b)$ follows from $U_s=(K,V_s)$. It results using (\ref{UhilfsvariableMIMO1}) that
 {\begin{align}
 		n[I(U_s;X_{J})-I(U_s;Y_{s,J})]
 		&=I(K;X^{n})-I(K;Y_s^{n}) \nonumber\\
 		&=H(K)-I(K;Y_s^{n})\nonumber \\ 
 		&=H(K|Y_s^n) 
 		\nonumber \\
            &=H(K|f(X^n),Y_s^{n})+ I(f(X^n);K|Y_s^n) \nonumber \\
            &\leq H(K|L_s)+ H(f(X^n)|Y_s^n) \nonumber \\
            &\overset{(a)}{\leq}  1+\mbb P\left[ K\neq L_s\right]\log\lvert \mc K \rvert +\log \lVert f \rVert \nonumber \\
            &\overset{(b)}{\leq}   1+n\alpha c +\log \lVert f \rVert  \nonumber \\
            &\leq 1+n\alpha c +n R, \nonumber
 		\end{align}}

   where $(a)$ follows from Fano's inequality and $(b)$ follows from \eqref{errorcorrelated1} and \eqref{cardinalitycorrelated1}.
   Thus, we obtain
   \begin{align}
       I(U_s;X_{J})-I(U_s;Y_{s,J})\leq R+\zeta(n,\alpha),
   \nonumber \end{align}
   where $\zeta(n,\alpha)=\frac{1}{n}+\alpha c.$
Since the joint distribution of $X_{J}$ and $Y_{s,J}$ is equal to $P_{XY_s},$ it follows that $\frac{H(K)}{n}$ is upper-bounded by $I(U_s;X)$ subject to $I(U_s;X)-I(U_s;Y_s) \leq R + \zeta(n,\alpha)$ with $U_s$ satisfying $U_s \circlearrow{X} \circlearrow{Y_s}.$ As a result, it follows using \eqref{ratecorrelated} that any achievable compound CR rate satisfies for every $s \in \mc S,$
\begin{align}
H <\underset{ \substack{U_s \\{\substack{U_s \circlearrow{X} \circlearrow{Y_s}\\ I(U_s;X)-I(U_s;Y_s) \leq R+\zeta(n,\alpha)}}}}{\max} I(U_s;X)+\delta.
 \label{righthandsideconverse}
\end{align}
By taking the limit when $n$ tends to infinity and then the infinimum over all $\alpha>0,\delta>0,$ of the right-hand side of \eqref{righthandsideconverse}, it follows that for every $s \in \mc S,$
\begin{align}
&H \leq \underset{ \substack{U_s \\{\substack{U_s \circlearrow{X} \circlearrow{Y_s}\\ I(U_s;X)-I(U_s;Y) \leq R}}}}{\max} I(U_s;X). \nonumber
\end{align}
This yields
\begin{align}
    H\leq \underset{s\in \mc S}{\min}\underset{ \substack{U_s \\{\substack{U_s \circlearrow{X} \circlearrow{Y_s}\\ I(U_s;X)-I(U_s;Y) \leq R}}}}{\max} I(U_s;X). 
\nonumber \end{align}
This completes the proof of the upper-bound on the compound CR capacity.

\section{Proof of Corollary \ref{corollary1}}
\label{specialcase1}
From Theorem \ref{compound CRcapacity}, we know that
    \begin{align}
        \underset{ \substack{U \\{\substack{\forall s \in \mc S: U\circlearrow{X} \circlearrow{Y_s}\\ I(U;X)-\underset{s\in \mc S}{\min}I(U;Y_s) \leq R}}}}{\max} I(U;X) \leq \underset{s\in \mc S}{\min} \underset{ \substack{U_s \\{\substack{U_s \circlearrow{X} \circlearrow{Y_s}\\ I(U_s;X)-I(U_s;Y_s) \leq R}}}}{\max} I(U_s;X)\leq H(X). \label{rewriteinequalities}
    \end{align}
  One can choose $U^\star=X$  such that for every $s\in \mc S,$ $U^\star\circlearrow{X}\circlearrow{Y_s}$ forms a Markov chain, such that $I(U^\star;X)=H(X)$ and that 
    \begin{align}
        I(U^\star;X)-\underset{s\in \mc S}{\min}\ I(U^\star;Y_{s}) &= H(X)-\underset{s\in \mc S}{\min} \ I(X;Y_{s}) \nonumber \\
        &=\underset{s\in \mc S}{\max} \ H(X|Y_s) \nonumber \\
        &\leq R.
    \end{align}
    Thus, we obtain
    \begin{align}
        \underset{ \substack{U \\{\substack{\forall s \in \mc S: U\circlearrow{X} \circlearrow{Y_s}\\ I(U;X)\leq R}}}}{\max} I(U;X)=I(U^\star;X)=H(X). \label{implication2}
    \end{align}
 \eqref{rewriteinequalities} and \eqref{implication2} imply that
 \begin{align}
        \underset{ \substack{U \\{\substack{\forall s \in \mc S: U\circlearrow{X} \circlearrow{Y_s}\\ I(U;X)-\underset{s\in \mc S}{\min}I(U;Y_s) \leq R}}}}{\max} I(U;X)= \underset{s\in \mc S}{\min} \underset{ \substack{U_s \\{\substack{U_s \circlearrow{X} \circlearrow{Y_s}\\ I(U_s;X)-I(U_s;Y_s) \leq R}}}}{\max} I(U_s;X)= H(X). 
    \nonumber \end{align}
This proves Corollary \ref{corollary1}.
\section{Proof of Corollary \ref{corollary2}}
\label{specialcase2}
We have 
    \begin{align}
\underset{ \substack{U \\{\substack{\forall s \in \mc S: U\circlearrow{X} \circlearrow{Y_s}\\ I(U;X)-\underset{s\in \mc S}{\min}I(U;Y_s) \leq R}}}}{\max} I(U;X) &\overset{(a)}{=}\underset{ \substack{U \\{\substack{\forall s \in \mc S: U\circlearrow{X}\circlearrow{Y_s}\circlearrow{Y_{s^\prime}}\\ I(U;X)-\underset{s\in \mc S}{\min}I(U;Y_s) \leq R}}}}{\max} I(U;X) \nonumber \\
&\overset{(b)}{=}\underset{ \substack{U \\{\substack{\forall s \in \mc S: U\circlearrow{X} \circlearrow{Y_s}\circlearrow{Y_{s^\prime}}\\ I(U;X)-I(U;Y_{s^\prime}) \leq R}}}}{\max} I(U;X) \nonumber \\
&= \underset{ \substack{U_{s^\prime} \\{\substack{U_{s^\prime}\circlearrow{X} \circlearrow{Y_{s^\prime}}\\ I(U_{s^\prime};X)-I(U_{s^\prime};Y_{s^\prime}) \leq R}}}}{\max} I(U_{s^\prime};X).\nonumber \\
&\geq\underset{s\in \mc S}{\min} \underset{ \substack{U_s \\{\substack{U_s \circlearrow{X} \circlearrow{Y_s}\\ I(U_s;X)-I(U_s;Y_s) \leq R}}}}{\max} I(U_s;X). \label{part1}
\end{align}
where $(a)$ follows because for every $s\in \mc S,$ $X\circlearrow{Y_{s}}\circlearrow{Y_{s^\prime}}$ forms a Markov chain and $(b)$ follows because every $s\in \mc S: I(U;Y_s)\geq  I(U;Y_{s^{\prime}}).$

Now, from Theorem \ref{compound CRcapacity}, we know that
\begin{align}
    \underset{ \substack{U \\{\substack{\forall s \in \mc S: U\circlearrow{X} \circlearrow{Y_s}\\ I(U;X)-\underset{s\in \mc S}{\min}I(U;Y_s) \leq R}}}}{\max} I(U;X)\leq \underset{s\in \mc S}{\min} \underset{ \substack{U_s \\{\substack{U_s \circlearrow{X} \circlearrow{Y_s}\\ I(U_s;X)-I(U_s;Y_s) \leq R}}}}{\max} I(U_s;X). \label{part2}
\end{align}
\eqref{part1} and \eqref{part2} yield
\begin{align}
    \underset{ \substack{U \\{\substack{\forall s \in \mc S: U\circlearrow{X} \circlearrow{Y_s}\\ I(U;X)-\underset{s\in \mc S}{\min}I(U;Y_s) \leq R}}}}{\max} I(U;X)= \underset{s\in \mc S}{\min} \underset{ \substack{U_s \\{\substack{U_s \circlearrow{X} \circlearrow{Y_s}\\ I(U_s;X)-I(U_s;Y_s) \leq R}}}}{\max} I(U_s;X). \nonumber
\end{align}
This proves Corollary \ref{corollary2}.


\section{Conclusion}
\label{conclusion}
In this paper, we addressed the problem of generating common randomness (CR) from finite compound sources with one-way communication over rate-limited perfect channels. Both terminals are assumed to know the set of source states as well as their statistics. However, they lack knowledge of the actual realization of the source state. Additionally, we assume that only the source outputs observed by the terminal at the receiving end of the noiseless channel are state-dependent.
We derived a single-letter lower and upper bound on the compound CR capacity for the specified model. \color{black}Notably, the upper bound we derived is known to be achievable when the terminal at the transmitting end of the noiseless channel knows the actual state, a condition not assumed in our work.\color{black}\
Furthermore, we present two scenarios where the derived bounds are equal.
\section*{Acknowledgments}
\color{black}
The authors acknowledge the financial support by the Federal Ministry of Education and Research
of Germany (BMBF) in the programme of “Souverän. Digital. Vernetzt.”. Joint project 6G-life, project identification number: 16KISK002.
H. Boche and R. Ezzine were further supported in part by the BMBF within the national initiative on Post Shannon Communication (NewCom) under Grant 16KIS1003K. 
C.\ Deppe was further supported in part by the BMBF within the national initiative on Post Shannon Communication (NewCom) under Grant 16KIS1005. C. Deppe was also supported by the German Research Foundation (DFG) within the project DE1915/2-1. M. Wiese was supported by the Bavarian Ministry of
Economic Affairs, Regional Development and Energy as part of the project 6G
Future Lab Bavaria.
\color{black}
\appendix
Recall that
\begin{align}
    \mathcal{A}_{\mbf M}= ``(X^n,Y_s^{n})\in  \mc E_{1}^{s}(\mbf M) \ \text{for all} \ s \in \mc S" \nonumber
\end{align}
and that
\begin{align}
    \mathcal{B}_{\mbf M}=``\exists s\in \mc S: (X^n,Y_s^{n})\in  \mc E_{2}^{s}(\mbf M)" \nonumber 
\end{align}
where for any $s\in \mc S$
\begin{align}
    \mc E_{1}^{s}(\mbf M)&=\{(x^{n},y^{n})\in \mc X^n\times\mc Y^n:\ (\Phi_{\mbf M}(x^{n}),x^{n},y^{n}) \notin \mathcal{T}_{\sigma_3}^{n} (P_{U,X,Y_s})\} \nonumber
\end{align} and 
\begin{align}
    &\mc E_{2}^{s}(\mbf M) \nonumber \\
    &=\Big\{(x^{n},y^{n})\in \mc X^n \times \mc Y^n: (x^{n},y^{n}) \in {\mc E_{1}^{s}(\mbf M)}^{c} \ \text{with} \ \bs{U}_{i,j}=\Phi_{\mbf M}(x^{n}) \nonumber \\   & \ \ \ \  \ \text{and such that} \ \exists \ \bs{U}_{i,\ell}\neq\bs{U}_{i,j} \ \text{where} \ (\bs{U}_{i,\ell},y^n)\in \mc \mc T_{\sigma_2}^{n}(P_{U,Y_s}) (\text{with the same first index} \ i)
\Big\}.\nonumber
\end{align}

We are going to show that for some $\zeta(n)>0$
\begin{align}
   \color{black} \mathbb{E}_{\mbf M}\left[ \mbb P\left[\mathcal{A}_{\mbf M}|\mbf M\right]+\mbb P\left[\mathcal{B}_{\mbf M}|\mbf M\right]\right]\color{black}\leq \zeta(n).
 \nonumber
\end{align} 
It holds that
\begin{align}
\color{black}\mathbb{E}_{\mbf M}\left[ \mbb P\left[\mathcal{A}_{\mbf M}|\mbf M\right]+\mbb P\left[\mathcal{B}_{\mbf M}|\mbf M\right]\right]= \mathbb{E}_{\mbf M}\left[ \mbb P\left[\mathcal{A}_{\mbf M}|\mbf M\right]\right]+\mathbb{E}_{\mbf M}\left[\mbb P\left[\mathcal{B}_{\mbf M}|\mbf M\right]\right]. \color{black} \label{summeans}
\end{align}
On the one hand, we have
\begin{align}
  \color{black}\mathbb{E}_{\mbf M}\left[ \mbb P\left[\mathcal{A}_{\mbf M}|\mbf M\right]\right]=\sum_{\mbf m}\mbb P\left[\mbf M=\mbf m\right] \mbb P\left[\mathcal{A}_{\mbf m}\right].\color{black} \label{boundzwisch3}
\end{align}
Now, for any s $\in \mc S$
\begin{align}
    \mbb P\left[\mathcal{A}_{\mbf m}\right] \leq  \mbb P\left[ (\Phi_{\mbf m}(X^{n}),X^{n},Y_s^{n}) \notin \mathcal{T}_{\sigma_3}^{n} (P_{U,X,Y_s})   \right]. 
    \label{boundzwisch1}
\end{align}
It holds that
\begin{align}
 \mbb P\left[ (\Phi_{\mbf m}(X^{n}),X^{n},Y_s^{n}) \notin \mathcal{T}_{\sigma_3}^{n} (P_{U,X,Y_s})   \right]=1- \mbb P\left[ (\Phi_{\mbf m}(X^{n}),X^{n},Y_s^{n}) \in \mathcal{T}_{\sigma_3}^{n} (P_{U,X,Y_s})   \right]
\nonumber \end{align}
and that
\begin{align}
    &\mbb P\left[ (\Phi_{\mbf m}(X^{n}),X^{n},Y_s^{n}) \in \mathcal{T}_{\sigma_3}^{n} (P_{U,X,Y_s})   \right] \nonumber \\ 
    &\geq \sum_{x^n \in \mc T_{\sigma_1}^{n}(P_{X}):(\Phi_{\mbf m}(x^n),x^n)\in \mc T_{\sigma_2}^{n}(P_{U,X})} P_{X^n}(x^n) \mbb P\left[(\Phi_{\mbf m}(x^n),x^n,Y_s^n) \in \mc T_{\sigma_3^{n}}(P_{U,X,Y_s})|X^n=x^n  \right] \nonumber \\
    &\overset{(a)}{\geq} \left(1-\kappa_{\sigma_2,\sigma_3}(n,P_{U,X,Y_s})\right) \mbb P\left[(\Phi_{\mbf m}(X^n),X^n)\in \mathcal{T}_{\sigma_2}^{n}(P_{U,X})\right] \nonumber
\end{align}
where $(a)$ follows from the \textit{Markov Lemma} \cite{markovlemma} since $U\circlearrow{X}\circlearrow{Y_s}$ forms a Markov chain, where
\begin{align}
  \kappa_{\sigma_2,\sigma_3}(n,P_{U,X,Y_s})=2\lvert \mc U \rvert \lvert \mc X \rvert \lvert \mc Y \rvert \exp\left(-2(1-\sigma_2)\left(\frac{\sigma_3 -\sigma_2}{1+\sigma_2} \right)^2 c_{U,X,Y_s}^2 n\right)\nonumber
\end{align}
with
\begin{align}
    c_{U,X,Y_s}= \underset{(u,x,y_s)\in \mc U \times \mc X \times \mc Y} {\min}{ P_{U,X,Y_s}(u,x,y_s)}.
\nonumber \end{align}

Let   $ \kappa_{\sigma_2,\sigma_3}(n,P_{U,X,Y_s^\star})=\underset{s\in \mc S}{\max} \ \kappa_{\sigma_2,\sigma_3}(n,P_{U,X,Y_s}).$

It follows that
\begin{align}
      &\mbb P\left[ (\Phi_{\mbf m}(X^{n}),X^{n},Y_s^{n}) \in \mathcal{T}_{\sigma_3}^{n} (P_{U,X,Y_s})   \right]\geq   \left(1-\kappa_{\sigma_2,\sigma_3}(n,P_{U,X,Y_s^\star})\right) \mbb P\left[(\Phi_{\mbf m}(X^n),X^n)\in \mathcal{T}_{\sigma_2}^{n}(P_{U,X}) \right]
\nonumber \end{align}
Therefore, we have
\begin{align}
    \mbb P\left[ (\Phi_{\mbf m}(X^{n}),X^{n},Y_s^{n}) \notin \mathcal{T}_{\sigma_3}^{n} (P_{U,X,Y_s})   \right] \leq 1- \left(1-\kappa_{\sigma_2,\sigma_3}(n,P_{U,X,Y_s^\star})\right) \mbb P\left[(\Phi_{\mbf m}(X^n),X^n)\in \mathcal{T}_{\sigma_2}^{n}(P_{U,X}) \right]
\nonumber \end{align}
As a result, \eqref{boundzwisch1} yields
\begin{align}
   \mbb P\left[\mathcal{A}_{\mbf m}\right]\leq  1- \left(1-\kappa_{\sigma_2,\sigma_3}(n,P_{U,X,Y_s^\star})\right) \mbb P\left[(\Phi_{\mbf m}(X^n),X^n)\in \mathcal{T}_{\sigma_2}^{n}(P_{U,X}) \right]
\nonumber \end{align}

This implies using \eqref{boundzwisch3} that
\begin{align}
 \mathbb{E}_{\mbf M}\left[ \mbb P\left[\mathcal{A}_{\mbf M}|\mbf M\right]\right] 
 \leq  1- \left(1-\kappa_{\sigma_2,\sigma_3}(n,P_{U,X,Y_s^\star})\right) \mbb E_{\mbf M}\left[\mbb P_{X^n}\left[(\Phi_{\mbf M}(X^n),X^n)\in \mathcal{T}_{\sigma_2}^{n}(P_{U,X})|\mbf M \right]\right]
\label{boundzwisch5} \end{align}
Now, we have
\begin{align}
  &\mbb E_{\mbf M}\left[\mbb P_{X^n}\left[(\Phi_{\mbf M}(X^n),X^n)\in \mathcal{T}_{\sigma_2}^{n}(P_{U,X})|\mbf M \right]\right] \nonumber \\
  &= \mbb E_{X^n}\left[\mbb P_{\mbf M}\left[(\Phi_{\mbf M}(X^n),X^n)\in \mathcal{T}_{\sigma_2}^{n}(P_{U,X}) \right]\right] \nonumber \\
  &\geq \sum_{x^n \in \mc T_{\sigma_1}^{n}(P_X)} P_{X^n}(x^n) \mbb P_{\mbf M}\left[(x^n,\Phi_{\mbf M}(x^n))\in \mc T_{\sigma_2}^{n}(P_{U,X})|X^n=x^n   \right] \nonumber \\
  &=\sum_{x^n \in \mc T_{\sigma_1}^{n}(P_X)} P_{X^n}(x^n) \mbb P_{\mbf M}\left[(x^n,\Phi_{\mbf M}(x^n))\in \mc T_{\sigma_2}^{n}(P_{U,X})   \right], \nonumber \\
\nonumber \end{align}
where we used that $\mbf M$ and $X^n$ are independent.
Now, for every $x^n \in \mc T_{\sigma_1^n}(P_X),$ we have
\begin{align}
    \mbb P_{\mbf M}\left[(x^n,\Phi_{\mbf M}(x^n))\in \mc T_{\sigma_2}^{n}(P_{U,X})   \right]&= 1-\mbb P\left[ \cap_{i,j} (\bs{U}_{i,j},x^n) \notin \mc T_{\sigma_2}^{n} (P_{U,X})\right] \nonumber \\
    &=1-\prod_{i,j} \mbb P\left[ (\bs{U}_{i,j},x^n) \notin \mc T_{\sigma_2}^{n} (P_{U,X})\right] \nonumber \\
    &=1-\prod_{i,j} \left[1-\mbb P\left[ (\bs{U}_{i,j},x^n) \in \mc T_{\sigma_2}^{n} (P_{U,X})\right]\right] \nonumber \\
    &\geq 1-\left[1-\left(1-\kappa_{\sigma_1,\sigma_2}(P_{U,X},n)  \right) 2^{-n\left[I(U;X)+2\sigma_2 H(U)   \right]}    \right]^{2^{n\left(I(U;X)+\mu\right)}} \nonumber \\
    &\geq 1-\exp\left( -2^{n\left(I(U;X)+\mu\right)}\left(1-\kappa_{\sigma_1,\sigma_2}(P_{U,X},n)  \right) 2^{-n\left[I(U;X)+2\sigma_2 H(U)   \right]}       \right) \nonumber \\
    &=1-\exp\left(-\left(1-\kappa_{\sigma_1,\sigma_2}(P_{U,X},n)  \right)2^{n\left[\mu-2\sigma_2H(U) \right]}  \right),
\nonumber \end{align}
where we used that $$ \mbb P\left[ (\bs{U}_{i,j},x^n) \in \mc T_{\sigma_2}^{n} (P_{U,X})\right] \geq \left(1-\kappa_{\sigma_1,\sigma_2}(P_{U,X},n)  \right) 2^{-n\left[I(U;X)+2\sigma_2 H(U)   \right]}$$ and that $(1-x)^m\leq \exp(-mx)$ and where
\begin{align}
    \kappa_{\sigma_1,\sigma_2}(P_{U,X},n)=2\lvert \mc U \rvert \lvert \mc X \rvert \exp \left(-2n(1-\sigma_1)\left(\frac{\sigma_2 -\sigma_1}{1+\sigma_1} \right)^2 c_{U,X}^2 n   \right)
\nonumber \end{align}
with $c_{U,X}(P_{U,X},n)=\underset{(u,x)\in \mc U\times \mc X}{\min}P_{U,X}(u,x).$

As a result, we have
\begin{align}
&\mbb E_{\mbf M}\left[\mbb P_{X^n}\left[(\Phi_{\mbf M}(X^n),X^n)\in \mathcal{T}_{\sigma_2}^{n}(P_{U,X}) \right]\right] \nonumber \\
&\geq \left[ 1-\exp\left(-\left(1-\kappa_{\sigma_1,\sigma_2}(P_{U,X},n)  \right)2^{n\left[\mu-2\sigma_2H(U) \right]}  \right) \right]\mbb P\left[ X^n \in \mc T_{\sigma_1}^{n}(P_X)  \right] \nonumber \\
&\geq \left[ 1-\exp\left(-\left(1-\kappa_{\sigma_1,\sigma_2}(P_{U,X},n)  \right)2^{n\left[\mu-2\sigma_2H(U) \right]}  \right) \right] \left[1-\kappa_{\sigma_1}(P_X,n)  \right],
\nonumber \end{align}
where $\kappa_{\sigma_1}(P_X,n)=2\lvert \mc X\rvert \exp\left(-2n\sigma_1 c_{X}^2  \right)$ with $c_{X}=\underset{x\in \mc X}{\min} P_{X}(x).$

Therefore, \eqref{boundzwisch5} yields
 \begin{align}
     & \color{black}\mathbb{E}_{\mbf M}\left[ \mbb P\left[\mathcal{A}_{\mbf M}|\mbf M\right]\right]\color{black}\nonumber \\     
      &\leq 1- \left(1-\kappa_{\sigma_2,\sigma_3}(n,P_{U,X,Y_s^\star})\right) \left[ 1-\exp\left(-\left(1-\kappa_{\sigma_1,\sigma_2}(P_{U,X},n)  \right)2^{n\left[\mu-2\sigma_2H(U) \right]}  \right) \right] \left[1-\kappa_{\sigma_1}(P_X,n)  \right]\nonumber \\
      &\color{black}\leq  \kappa_{\sigma_1}(P_X,n)+ \exp\left(-\left(1-\kappa_{\sigma_1,\sigma_2}(P_{U,X},n)  \right)2^{n\left[\mu-2\sigma_2H(U) \right]}  \right)+  \kappa_{\sigma_2,\sigma_3}(n,P_{U,X,Y_s^\star}) \nonumber \\ 
      &\quad \color{black}+     \kappa_{\sigma_1}(P_X,n)\exp\left(-\left(1-\kappa_{\sigma_1,\sigma_2}(P_{U,X},n)  \right)2^{n\left[\mu-2\sigma_2H(U) \right]}  \right)\kappa_{\sigma_2,\sigma_3}(n,P_{U,X,Y_s^\star}) \color{black}  
 \label{firstbound} \end{align} 

 On the other hand, we have
 \begin{align}
&\color{black}\mathbb{E}_{\mbf M}\left[ \mbb P\left[\mathcal{B}_{\mbf M}|\mbf M\right]\right] \nonumber \\
  &\color{black}\leq \sum_{s\in \mc S} \sum_{i=1}^{N_1}\sum_{j=1}^{N_2}\sum_{\ell=1,\ell\neq j}^{N_2}\nonumber \\  &\quad \quad \quad \quad \quad \color{black}\mathbb{E}_{\bs{U}_{i,j},\bs{U}_{i,\ell}}\left[ \mbb P\left[ (\bs{U}_{i,j},X^n,Y_{s}^{n}) \in \mc T_{\sigma_3}^{n}(P_{U,X,Y_s}),(\bs{U}_{i,j},X^n)\in \mc T_{\sigma_2}^{n}(P_{U,X}), (\bs{U}_{i,\ell},Y_s^{n})\in \mc T_{\sigma_2}^{n}(P_{U,Y_s})|\bs{U}_{i,j},\bs{U}_{i,\ell}\right] \right] \nonumber \\ \color{black}
  &\color{black}= \sum_{s\in \mc S} \sum_{i=1}^{N_1}\sum_{j=1}^{N_2}\sum_{\ell=1,\ell\neq j}^{N_2} \nonumber \\
  &\quad \quad \quad \quad \quad \color{black}\mathbb{E}_{X^n,Y_s^n}\left[ \mbb P_{\bs{U}_{i,j},\bs{U}_{i,\ell}}\left[ (\bs{U}_{i,j},X^n,Y_{s}^{n}) \in \mc T_{\sigma_3}^{n}(P_{U,X,Y_s}),(\bs{U}_{i,j},X^n)\in \mc T_{\sigma_2}^{n}(P_{U,X}), (\bs{U}_{i,\ell},Y_s^{n})\in \mc T_{\sigma_2}^{n}(P_{U,Y_s})\right] \right] \nonumber \\
  &=\sum_{s\in \mc S} \sum_{i=1}^{N_1}\sum_{j=1}^{N_2}\sum_{\ell=1,\ell\neq j}^{N_2} \sum_{(x^n,y_s^n)\in \mc T_{\sigma_1}^{n}(P_X)\times \mc T_{\sigma_1}^{n}(P_{Y_{s}})} \nonumber \\ 
  &\quad \quad \quad \quad \quad P_{X^n,Y_{s}^n}(x^n,y_{s}^n)  \mbb P\left[ (\bs{U}_{i,j},x^n,y_s^{n}) \in \mc T_{\sigma_3}^{n}(P_{U,X,Y_s}),(\bs{U}_{i,j},x^n)\in \mc T_{\sigma_2}^{n}(P_{U,X}), (\bs{U}_{i,\ell},y_s^{n})\in \mc T_{\sigma_2}^{n}(P_{U,Y_s})\right]  \nonumber \\
  &\leq \sum_{s\in \mc S} \sum_{i=1}^{N_1}\sum_{j=1}^{N_2}\sum_{\ell=1,\ell\neq j}^{N_2} \sum_{(x^n,y_s^n)\in \mc T_{\sigma_1}^{n}(P_X)\times \mc T_{\sigma_1}^{n}(P_{Y_{s}})} P_{X^n,Y_{s}^n}(x^n,y_{s}^n) \mbb P\left[ (\bs{U}_{i,j},x^n)\in \mc T_{\sigma_2}^{n}(P_{U,X}), (\bs{U}_{i,\ell},y_s^{n})\in \mc T_{\sigma_2}^{n}(P_{U,Y_s})\right]  \nonumber \\
  &=\sum_{s\in \mc S} \sum_{i=1}^{N_1}\sum_{j=1}^{N_2}\sum_{\ell=1,\ell\neq j}^{N_2} \sum_{(x^n,y_s^n)\in \mc T_{\sigma_1}^{n}(P_X)\times \mc T_{\sigma_1}^{n}(P_{Y_{s}})} P_{X^n,Y_{s}^n}(x^n,y_{s}^n) \mbb P\left[ (\bs{U}_{i,j},x^n)\in \mc T_{\sigma_2}^{n}(P_{U,X})\right] \mbb P\left[(\bs{U}_{i,\ell},y_s^{n})\in \mc T_{\sigma_2}^{n}(P_{U,Y_s})\right] 
\nonumber \end{align} 

Now, for every $x^n \in \mc T_{\sigma_1}^{n}(P_X),$ it holds that
\begin{align}
    \mbb P\left[ (\bs{U}_{i,j},x^n)\in \mc T_{\sigma_2}^{n}(P_{U,X})\right] \leq 2^{-n\left[I(U;X)-2\sigma_2 H(U)  \right]}
\nonumber \end{align}
and for every $y_s^n \in \mc T_{\sigma_1}^{n}(P_{Y_s}),$ it holds that
\begin{align}
    \mbb P\left[ (\bs{U}_{i,\ell},y_s^n)\in \mc T_{\sigma_2}^{n}(P_{U,Y_s})\right] \leq 2^{-n\left[I(U;Y_s)-2\sigma_2 H(U)  \right]}
\nonumber \end{align}
Hence, it follows that
\begin{align}
    &\color{black}\mathbb{E}_{\mbf M}\left[ \mbb P\left[\mathcal{B}_{\mbf M}|\mbf M\right]\right]\color{black} \nonumber \\
    &\leq \lvert \mc S \rvert 2^{n\left(\underset{s\in \mc S}{\min}I(U;Y_s)-2\mu\right)}2^{n\left(I(U;X)+\mu\right)} 2^{-n\left[I(U;X)-2\sigma_2 H(U)  \right]}2^{-n\left[I(U;Y_s)-2\sigma_2 H(U)  \right]} \nonumber \\
    &\color{black}\leq \lvert \mc S\rvert 2^{n \left[-\mu+4\sigma_2 H(U)\right]}
\label{secondbound} \end{align}
As a result, \eqref{summeans}, \eqref{firstbound} and \eqref{secondbound} yield
\begin{align}
      \color{black}\mathbb{E}_{\mbf M}\left[ \mbb P\left[\mathcal{A}_{\mbf M}|\mbf M\right]+\mbb P\left[\mathcal{B}_{\mbf M}|\mbf M\right]\right] \leq \zeta(n) \nonumber
\end{align}
where
\begin{align}
    \zeta(n)&=\kappa_{\sigma_1}(P_X,n)+ \exp\left(-\left(1-\kappa_{\sigma_1,\sigma_2}(P_{U,X},n)  \right)2^{n\left[\mu-2\sigma_2H(U) \right]}  \right)+ \kappa_{\sigma_2,\sigma_3}(n,P_{U,X,Y_s^\star}) \nonumber \\ 
      &\quad \color{black}+ \kappa_{\sigma_1}(P_X,n)\exp\left(-\left(1-\kappa_{\sigma_1,\sigma_2}(P_{U,X},n)  \right)2^{n\left[\mu-2\sigma_2H(U) \right]}  \right)\kappa_{\sigma_2,\sigma_3}(n,P_{U,X,Y_s^\star}) \nonumber \\
      &\quad \quad +\lvert \mc S\rvert 2^{n \left[-\mu+4\sigma_2 H(U)\right]}. \nonumber
\end{align}
and where $4\sigma_2 H(U)< \mu <5\sigma_2H(U).$
Clearly, $\underset{n\rightarrow\infty}{\lim} \zeta(n)=0.$

\clearpage
\newpage
\end{document}